\documentclass[12pt]{article}
\usepackage{amsmath}
\usepackage{amssymb}

\newcommand{\C}{\mathbf{C}}

\newcommand{\F}{\mathbf{F}}
\newcommand{\Fop}{\mathbf{F}\mathrm{(op)}}
\newcommand{\Fv}{\mathbf{F}(v)}
\newcommand{\I}{\mathbf{I}}

\newcommand{\boldg}{\mathbf{g}}
\newcommand{\bfsigma}{\mathbf{\sigma}}
\newcommand{\X}{\mathbf{X}}
\newcommand{\bQ}{\mathbf{Q}}
\newcommand{\bU}{\mathbf{U}}
\newcommand{\bV}{\mathbf{V}}
\newcommand{\Q}{\mathbf{Q}}
\newcommand{\U}{\mathbf{U}}
\newcommand{\V}{\mathbf{V}}

\newcommand\diag{\mathrm{diag}}

\begin{document}

\title{Continuation of the Fermion-Number Operator
and the Puzzle of Families\footnote{To be presented at DPF 2000 at The Ohio State University, 
Columbus, OH, August 9--12, 2000. Download from the APS e-print archive at: http://publish.aps.org.eprint/. Paper number: aps2000mar19\underbar{~}003.}}
\author{Gerald L. Fitzpatrick}
\date{PRI Research and Development Corp.\\
12517 131 Ct.\  N.\ E.\\
Kirkland, WA 98034}

\maketitle

\begin{abstract}
An ``analytic continuation''
of a Hermitian matrix representing
the conventional fermion-number \emph{operator}, leads to a
new, and unconventional,
\emph{internal} description of \emph{quarks} and \emph{leptons}. This phenomenological
description, unlike the conventional standard-model description, is capable of explaining, 
among other things, why there are just \emph{three} families of quarks and leptons. These facts provide indirect evidence that  the analytic continuation in question somehow reflects physics at the Planck level where flavor degrees-of-freedom presumably originate.
\end{abstract}

\setcounter{section}{1}
\setcounter{subsection}{-1}

\subsection{Introduction and Background} 

Given the number of flavors of quarks and leptons, and an appropriate (renormalizable) lagrangian, the so-called ``accidental symmetries'' of the lagrangian [1] are known to
``explain'' the separate conservation of various (global) ``charges'' (e.g., lepton number, baryon number, strangeness, charm, truth, beauty, electron-, muon- and tau-numbers).
However,  there is nothing in such lagrangians, or their
associated accidental symmetries, that would explain quarks and leptons,
or tell us how many flavors of quarks and leptons to include. What is needed is a ``spectrum-generating'' mechanism for these
particles. 

The current ``consensus'' in particle physics, at least among string theorists, seems to be that fundamental-fermion flavor degrees-of-freedom arise at the Planck level [2, 3]---or possibly at $TeV$ energies in the event that  the extra dimensions are ``large'' [4]---in a theory of superstrings.
The purpose of the present paper is to identify an apparently different, but probably complementary, (phenomenological) spectrum-generating
mechanism that arises in what seems to be a most unlikely way. 

We have found that an ``analytic-continuation'' [5] of a Hermitian matrix $\Fop$ representing
the conventional fermion-number
\emph{operator}, from an \emph{external} spacetime and Hilbert-space setting
to a new \emph{internal} (real) non-Euclidean space---$\Fop$ is continued to a real, generally  non-diagonal matrix 
$\Fv$ involving a single \emph{real} parameter $v$---``automatically'' leads to a new description of fundamental fermions (quarks and leptons) in
which families are \emph{replicated} and there are just \emph{three} families. The fact that this happens,
suggests that there is some deep connection between the result of the aforementioned 
``analytic-continuation,'' namely $\Fv$, and Planck-level physics where flavor degrees-of-freedom, and family-replication, presumably originate.

The work presented here is based on a little-known book by the author [6] wherein many of the
consequences of using $\Fv$ to describe fundamental fermions, are worked out in
much greater detail than in the present paper. And, conversely, certain themes that were only alluded to in [6], such as the ``analytic-continuation'' $\Fop\to\Fv$, are discussed in much greater detail here. The author hopes that the following relatively brief ``overview'' and extension of [6] will serve to increase awareness of these new ideas among particle physicists.
   
\setcounter{section}{2}
\setcounter{subsection}{-1}

\subsection{The Conventional Fermion-Number Operator}

Consider the situation, presumably at some high energy, where we are dealing with ``free'' (isolated) leptons or ``analytically free'' quarks. Suppose we want to describe the scalar fermion-number carried by these particles. And, suppose further that the energies involved are not so high that quantum field-theory  (QFT) breaks down. Under these conditions the fermion-number can be represented by  a $U(1)$-type scalar ``charge'' [7, 8], namely, a charge associated with the (continuous)
group of  unitary matrices $U$ of order 1 known as $U(1)$. 

The
fermion-number operator, which can be represented by a Hermitian matrix $\Fop$, is said to generate
these so-called ``gauge'' (or phase) transformations, which in turn act on
fermion and antifermion quantum states in Hilbert space. That is,
given that $\alpha$ is a \emph{real} phase  one has
\begin{equation}\label{eqn1}
U=e^{i\alpha\Fop},
\end{equation}
and for infinitesimal transformations [i.e., $e^{i\,\delta\alpha\,\Fop}=
1+i\,\delta\alpha\,\Fop$] acting on single-particle (free or ``asymptotically free'')
fermion and antifermion states
$|p\rangle$ and $|\overline p\rangle$, respectively, one easily
establishes that (the fermion-number ``charges'' are  $f_m=-f_a=1$ for matter and $f_a$ for antimatter)
\begin{equation}\label{eqn2}
\begin{array}{rcl}
U|p\rangle & =  e^{i\,\delta\alpha\,f_m}|p\rangle \\
U|\overline p\rangle & = e^{i\,\delta\alpha\,f_a}|\overline p\rangle, 
\end{array}
\end{equation}
since, by definition, $\Fop$ obeys the eigenvalue equations
\begin{equation}\label{eqn3}
\begin{array}{rcl}
\Fop|p\rangle & = f_m|p\rangle \\
\Fop|\overline p\rangle & =f_a|\overline p\rangle.
\end{array}
\end{equation}

Finally, the assumption that the Hamiltonian  $H$ is invariant under $U$, namely
\begin{equation}\label{eqn4}
H=UHU^\dagger,
\end{equation}
ensures that $H$ and $\Fop$ commute
\begin{equation}\label{eqn5}
[\Fop, H]=0,
\end{equation}
as can be verified by differentiating $UHU^\dagger$ with respect to
$\alpha$. Hence, the total fermion-number (the number of \emph{fermions} minus the
number of \emph{antifermions}) is a constant of the motion.

\subsection{Matrix representation of the fermion-number operator}

Because the matrix $\Fop$ involves just \emph{two} kinds of
quantum states (\ref{eqn3}),
namely $|p\rangle$ and $|\overline p\rangle$,  it can be expressed as a $2\times 2$ diagonal Hermitian matrix (see below), where one of the 
adjustable parameters $(\theta)$ is a \emph{fixed} constant (up to $2\pi$)
and the other $(\phi)$ is freely \emph{adjustable}. In particular, 
\begin{equation}\label{eqn6}
\Fop = \left.\left( \begin{array}{ll}
\cos\theta & \sin\theta e^{-i\phi} \\
\sin\theta e^{+i\phi} & -\cos\theta
\end{array}\right)\right|_{\cos\theta=1}=\sigma_z,
\end{equation}
where 
\begin{equation}\label{eqn7}
\sigma_z=\left(\begin{array}{cc}
f_m & 0 \\
0 & f_a\end{array}\right)
\end{equation}
is one of the familiar Pauli matrices.

This form for $\Fop$ is consistent with (\ref{eqn3}) 
where the normalization and orthogonality conditions, namely $\langle p|p\rangle = \langle\overline p|\overline p\rangle=1$ and $\langle p|\overline p\rangle = \langle\overline p|p\rangle =0$, respectively, directly yield
\begin{equation}\label{eqn8}
\Fop = \left( \begin{array}{ccc}
\langle p|\Fop|p\rangle & , & \langle p|\Fop|\overline p\rangle
\\
\langle\overline p|\Fop|p\rangle & , & \langle\overline
p|\Fop|\overline p\rangle
\end{array}\right)=\sigma_z.
\end{equation}
Note that owing to (\ref{eqn7}) and (\ref{eqn8}), $\cos\theta<1$ in (\ref{eqn6}) is excluded.
Here it should also be noted that $tr\Fop=f_m+f_a=0$, $det\Fop=f_m\cdot f_a=-1$, and $\mathbf{F}^2\mathrm{(op)}=\I_2$ is the $2\times 2$ identity matrix.

\setcounter{section}{3}
\setcounter{subsection}{-1}

\subsection{The Continuation From F(op) to F($v$)}

Now perform an ``analytic continuation'' [5] on $\Fop$, namely $\Fop\to\Fv$, which maintains $\Fv$ \emph{real} and $\cos\theta\ge 1$. This can only be accomplished by continuing $\theta$ from a  \emph{real} to an \emph{imaginary} number, and by maintaining $e^{-i\phi}$  \emph{imaginary}. In particular, to maintain $\Fv$ real, we must make the replacements $\theta\to iv$ and $e^{-i\phi}\to \mp
i$, where $v$ is a \emph{real} number. Then
\begin{eqnarray}
\Fv & = & \left.\left(\begin{array}{ll}
\cos\theta & \sin\theta e^{-i\phi} \\
\sin\theta e^{+i\phi} & -\cos\theta\end{array}\right)\right|_{\theta=iv, e^{-i\phi}=\mp
i}\\ \label{eqn9}
&\hbox{or}& \nonumber \\
\Fv & = & \left(\begin{array}{ll}
\cosh v & \pm \sinh v \\
\mp\sinh v & -\cosh v \end{array}\right), \label{eqn10}
\end{eqnarray}
where, just as for $\Fop$ the eigenvalues of $\Fv$ are $f_m$ and $f_a$, and so we have $tr\Fv=f_m+f_a=0$, $det\Fv=f_m\cdot f_a=-1$, and $\F^2(v)=\I_2$.

In [6, p.\ 50 and 54] it is shown that only the upper signs in (\ref{eqn10}) have physical significance and $v\ge 0$.
And, just what is the physical significance of $\Fv$?

Because the continuation ``connects'' $\Fop$ and $\Fv$, it is natural to assume that both $\Fop$ and $\Fv$  describe, or represent, aspects of the fermion number (i.e., the matter-antimatter ``degree-of-freedom''). However, unlike $\Fop$, $\Fv$ will be shown to describe additional ``degrees-of-freedom'' such as the ``up''-``down'' and quark-lepton ``degrees-of-freedom.''
Moreover, it is abundantly clear from (\ref{eqn10}) that the the generally non-Hermitian (when $v\ne 0$) matrix $\Fv$---unlike the Hermitian matrix $\Fop$ in (\ref{eqn8}), which acts on Hilbert space---does \emph{not} act on a Hilbert space in an \emph{external} spacetime setting
[9].

\subsection{A new internal non-Euclidean space}

When the matrix $\Fv$ acts on a real column-vector $\{a, b\}$, it leaves the
quadratic form
$a^2-b^2$, invariant.  Therefore, the 2-space metric is non-Euclidean or
``Lorentzian'', and can be represented by the matrix
\begin{equation}\label{eqn11}
\boldg = \left(
\begin{array}{cc}
1 & 0 \\
0 & -1 \end{array}\right).
\end{equation}
Given this metric, the \emph{scalar product} of two real vectors assumes the
form
\begin{equation}\label{eqn12}
(a,b)\{e,f\}=ae-bf.
\end{equation}
Similarly,  the \emph{square} of a real vector is given by
\begin{equation}\label{eqn13}
(a,b)\{a,b\}=a^2-b^2.
\end{equation}

Here $(\;,\;)$ is a \emph{row} vector while $\{\;,\;\}$ is a (conformable)
\emph
{column} vector.
Now let us demonstrate that these  scalar-products  transform like \emph{charges} instead of \emph{probabilities} as they would if we were still dealing with a Hilbert space. 

\subsection{Charge conjugation}

The matrix $\X$, where $\X = \X^{-1}$ or $\X^2 = \I_2$ , given by
\begin{equation}
\X = \left( \begin{array}{cc}
0 & 1/d \\
d & 0 \end{array}\right),
\end{equation}
transforms $\Fv$ to is $\C$-reversed counterpart.  In particular, given that
the
 components of 
$\Fv$, in diagonal form [i.e., $\Fv_{\diag}=\sigma_z$], transform like (global) ``charges" ($f_m$ and $f_a$), the general similarity
transformation must be such that 
\begin{equation}
\X\;\Fv\;\X^{-1}=-\Fv,
\end{equation}
for \emph{any} real $d$.  However, since $\X$ should convert a real vector $\{a,
b\}$
 to its 
\emph{orthogonal} antimatter-counterpart
\begin{equation}
\X\;\{a,b\}=\{b/d,ad\},
\end{equation}
we are forced to require  
\begin{equation}
(a,b)\{b/d,ad\}=0,
\end{equation}
which means that 
\begin{equation}
ab/d-abd=0,
\end{equation}
or $d^2 = 1$.  Therefore, we have two choices for $d$, namely, $d = \pm 1$.
Which one describes the 
physics of fundamental fermions?

Clearly, when  $d= \pm 1$,  the \emph{square} of any vector, namely, 
\begin{equation}
(a,b)\{a,b\}=a^2-b^2,
\end{equation}
changes signs (is $\C$-reversing) under $\X$.  However, in
general, the components of a vector $\{a, b\}$
do not change signs under $\X$, i.e., they are not $\C$-reversing.

For example, when $d = +1$ in (14), the components of real vectors such as
$\{a, b\}$ are \emph{not} $\C$-reversing
charge-like quantities, which means that we cannot represent $\C$-reversing
(Lorentz 4-scalar) charges such as
electric-charge, charm, isospin or beauty as \emph{components}
of such vectors when $d = +1$.  However,
when $d = -1$,  both the square of
$\{a, b\}$ \emph{and} its
components $a$ and $b$ change signs (are $\C$-reversing) under $\X$, and can be used to represent such charges.
Therefore, from an experimental
standpoint, $d = +1$ is excluded (see Ref. 6, p.\ 8).

We conclude that the matrix $\X$, which plays the role of
charge-conjugation  (with respect to these global 2-space charges only) in this non-Euclidean
2-space, is proportional to one of the familiar Pauli matrices, namely,
\begin{equation}
\X=-\bfsigma_x.
\end{equation}

The foregoing properties of the non-Euclidean charge-like scalars, leads to the following conjecture:
\medskip

\noindent\emph{The global {\rm(}flavor-defining{\rm)} charges associated with the aforementioned ``accidental'' symmetries {\rm(}see Section 1.0{\rm)}, and the global charges associated with the non-Euclidean 2-space, are {\rm(}essentially{\rm)} one and the same charges.}
\medskip

As shown in the next section if this conjecture is true it means, among other things, that simultaneous fundamental-fermion flavor eigenstates can be partially specified using an appropriate (mutually-commuting) combination of these (global) 2-space charges [10, 11].

\setcounter{section}{4}
\setcounter{subsection}{-1}

\subsection{Representing Flavor Eigenstates and Flavor Doublets in the 2-Space}

The continuation from the 2-D Hilbert space to the 2-D non-Euclidean
``charge" space, turns out to mean that individual flavors of fundamental fermions 
can be partially  represented in an unconventional way by geometric objects (in the non-Euclidean 2-space) 
which \emph{differ} from a quantum state, but \emph{from which the quantum states 
can be inferred or effectively constructed}.  In particular, in the non-Euclidean 2-space, an object 
we call a ``vector triad" represents ``up"-`` down" type flavor doublets of
fundamental fermions---the ``up"-``down" type flavor dichotomy.  That is, 
the components of the vectors associated with a given vector-triad 
are \emph{observable} (Lorentz 4-scalar) ``charges," which can be 
used to (partially) define the two flavor-eigenstates in a flavor doublet [10, 11].

\subsection{Flavor eigenstates}

As demonstrated in detail in 
[6, pp.\ 16--18], given the charge-like ($\C$-reversing) \emph{observables} associated
with the description involving 
$\Fv$, namely, the real $\C$-reversing scalar-components of various matrices
and
 vectors defined on 
the internal non-Euclidean 2-space, it is possible to write down flavor
eigenstates [10, 11].

What one does is to identify the mutually-commuting $\C$-reversing 
``charges" (call them $C_i$) or charge-like quantum numbers associated 
with a particular flavor, and then write the corresponding simultaneous 
flavor-eigenstate as
\begin{equation}
|C_1, C_2, C_3, \ldots, C_n\rangle.
\end{equation}
Here $C_1, C_2, C_3, \ldots, C_n$, are the ``good" charge-like quantum 
numbers (charges) associated with a particular flavor.  It happens that 
these \emph{observable}  real-numbers can be identified with quantum 
numbers such as \emph{electric charge, strangeness, charm}, the 
third-component of (global) \emph{isospin, truth and beauty} (See Ref.\ 6, p.\ 72).
To discover what charges describe a particular flavor we
must first identify the vector-triad associated with that flavor.

Now, each \emph{vector-triad} represents a flavor doublet, not just an individual
flavor.  That is,
vector-triads provide information on \emph{two} quantum states (simultaneous
flavor-eigenstates)
associated with flavor doublets.  Therefore, vector-triads represent both
individual flavors \emph{and}
flavor doublets.  Here we simply summarize
how it is the that non-Euclidean vector-triads represent both
\emph{individual} flavors and ``up"-``down" type
\emph{flavor-doublets}.

\subsection{Flavor doublets}

Consider the eigenvector (call it $\bQ$) of $\Fv$ for fundamental fermions [12].
Since the space on which $\Fv$
``acts" is two-dimensional, the \emph{observable} vector $\Q$ can
be ``resolved" into two (no more or less)
\emph{observable},  linearly-independent vectors, call them $\U$ and $\V$,
as
$\Q = \U + \V$ [13].  Now, because these
three vectors ($\Q$, $\U$, and $\V$) are \emph{simultaneous}
observables, it makes sense to speak of this ``triad"
of vectors as being a well defined geometric object, namely, a ``vector triad."

Recognizing that the components of $\Q$, $\U$ and $\V$ are $\C$-reversing
charge-like \emph{observables} we can
write these \emph{observable} ``charge" vectors as
\begin{eqnarray}
\Q & = & \{q_1, q_2\} \\
\U & = & \{u_1, u_2\} \\
\V & = & \{v_1, v_2\},
\end{eqnarray}
where $q_1$, $q_2$, $u_1$, $u_2$, $v_1$ and $v_2$ are the various
\emph{observable} ``charges" (e.g., $q_1$ and $q_2$ are found to be \emph{electric} charges). 
Given $\Q = \U + \V$, the non-Euclidean metric (11), and Eqs. (22) through
(24), we
 find the 
associated \emph{observable} quadratic-``charges"   
\begin{eqnarray}
\Q^2 & = & \U^2 + 2 \U\bullet\V + \V^2 \\
2\U\bullet\V & = & 2(u_1v_1-u_2v_2) \\
\U^2 & = & u^2_1 - u^2_2 \\
\V^2 & = & v^2_1-v^2_2.
\end{eqnarray}
Finally, using the foregoing charges, we can express the \emph{two} quantum
states (simultaneous 
flavor-eigenstates) associated with a \emph{single} vector-triad in the
form of ``ket"vectors as follows (Ref.\ 6, pp.\ 16--18)
\begin{equation}
\begin{array}{l}
|q_1, u_1, v_1, \Q^2, \U^2, 2\U\bullet \V, \V^2\rangle, \\
|q_2, u_2, v_2, \Q^2, \U^2, 2\U\bullet \V, \V^2\rangle.
\end{array}
\end{equation}
Here, the state $|q_1, u_1, v_1, \Q^2, \U^2, 2\U\bullet \V, \V^2\rangle$
 represents the ``up"-type flavor-eigenstate, and
$|q_2, u_2, v_2, \Q^2, \U^2, 2\U\bullet\V, \V^2\rangle$
 represents the corresponding ``down"- type flavor-eigenstate
in a flavor doublet of fundamental fermions [10, 11].

Up to this point in the discussion we have shown that the 2-space description naturally incorporates the matter-antimatter dichotomy and the ``up''-``down'' flavor dichotomy. Now let us demonstrate how quarks and leptons are incorporated in the new description.

\subsection{Distinguishing quarks and leptons}

Choosing the upper signs in (10), the matrix $\Fv$ becomes
\begin{equation}
\F(v) = \left( \begin{array}{cc}
\cosh v & \sinh v\\
-\sinh v & -\cosh v
\end{array}\right),
\end{equation}
where $v$ is a positive real number [6, p.\ 50 and 54].

As described in [6, pp.\ 52--55], the parameter $v$ distinguishes between \emph{quarks} and
\emph{leptons}.  In particular, the 
parameter $v$ is found to be \emph{quantized} and obeys the ``quantum condition":  
\begin{equation}
v=\ln M_c,
\end{equation}
where $M_c$ counts both the \emph{number} of fundamental fermions in a
strongly-bound composite 
fermion, and the \emph{strong-color multiplicity}.  That is, $M_c = 3$ for quarks
(strong-color triplets) and 
$M_c = 1$ for leptons (strong-color singlets).  

\subsubsection{Quark and lepton electric charges}

It is shown in (Ref.\ 6, pp.\ 52--55, and Ref.\ 12) that the quark and lepton electric charges are the 
``up''-``down'' components of the eigenvectors of the matrix $\Fv$. In particular, the quark charges are given by $(M_c=3)$
\begin{eqnarray}
q_1(f) & = & \frac{(M^2_c-1)}{2M_c(M_c-f)} = +\frac{2}{3}\hbox{ for }f=+1\hbox{ and }+\frac{1}{3}\hbox{ for }f=-1 \\
q_2(f) & = & q_1(f)-1.
\end{eqnarray}
Similarly, the lepton electric charges are given by $(M_c=1)$
\begin{eqnarray}
q'_1(f) & = & \frac{-(M_c^2-1)}{2M_c(M_c-f)} = -1\hbox{ for }f=+1\hbox{ and }0\hbox{ for }f=-1 \\
q'_2(f) & = & q'_1(f)+1.
\end{eqnarray}
In summary, 
$\Fv$ is found to provide an 
explanation for the quark-lepton ``dichotomy" of fundamental fermions in
addition to the matter-antimatter, and ``up"-`` down" type flavor-dichotomy.  

\subsection{Family replication and the number of families}

In [6, pp.\ 59--65] it is shown that flavor doublets (hence families) are \emph{replicated} and that there are only three families of quarks and leptons. We  refer the reader to [6] for a full and detailed account. Here we simply outline how this situation comes about.

By the definition of a linear-vector 2-space, a 2-vector such as $\Q$ can always be resolved into a pair (no more, or less) of linearly-independent vectors $\U$ and $\V$ as $\Q=\U+\V$ (see Sec.\ 4.2). And, since $\Q$ \emph{represents} a flavor doublet, so should $\U$ and $\V$ \emph{represent} this \emph{same} flavor doublet. But, if this is so, \emph{different vector-resolutions of $\Q$ 
{\rm(}i.e., different vector-triads{\rm)} should correspond to different flavor-doublets having the same $\Q$}. In other words, flavor doublets should be \emph{replicated}.

Since $\Q$ can be resolved (mathematically) in an infinite number of ways, we might suppose that there are an \emph{infinite} number of flavor doublets, and hence, families. But, because of various ``quantum constraints,'' it is possible to show that $\Q$ \emph{can be resolved in only three physically acceptable ways for $\Q$-vectors associated with either quarks or leptons}. In other words, there can be only six quark flavors and six lepton flavors, which leads to the 
(\emph{ex post facto})``prediction'' of three quark-lepton families.

\setcounter{section}{5}
\setcounter{subsection}{-1}

\subsection{Discussion}

It is important to understand that the new 2-space description of fundamental fermions (quarks and leptons) provides a distinction between these particles that goes beyond differences that can be explained by mass differences alone. For example, in the standard model the only difference between an electron and a muon is that they have different masses. Otherwise, these particles experience identical electroweak interactions. Moreover, as described in Section 1.0, the separate conservation of electron- and muon-numbers can be attributed to certain unavoidable ``accidental symmetries'' associated with the (renormalizable) lagrangian describing the (electroweak) interactions of these particles.

Taken at face value, these accidental symmetries would seem to imply that there are no internal ``wheels and gears'' that would distinguish an electron from a muon, for example. But, if the string theories are correct, these particles would be associated with different ``handles'' on the compactified space [see Ref.\ 3, Vol.\ 2, p.\ 408], and so would be different in this \emph{additional} sense. Likewise, in the present non-Euclidean
2-space description, a variety of (global) 2-scalars, which are only \emph{indirectly} related to the  accidental symmetries of the lagrangian, serve to provide a (further) distinction between particles such as the electron and muon (see also the conjecture in Sec.\ 3.2).

A probable experimental signal of such ``internal'' differences is to be found in the recent observations at the Super Kamiokande of bi-maximal neutrino mixing [14]. Models which begin by positing a neutrino mass-matrix and associated mixing-parameters, such as the three-generation model proposed by Georgi and Glashow [15], do a good job of describing the observations. However, bi-maximal mixing may have a deeper explanation in terms of internal topological-differences (in the non-Euclidean 2-space) between $\nu_e$, and $\nu_\mu$ or $\nu_\tau$ neutrinos.

With respect to the internal transformation $\Fv$, the topology of the non-Euclidean ``vector triad'' (see Sec.\ 4.2) representing the $\nu_e$ ($\nu_\mu$ or $\nu_\tau$), is found to be that of a cylinder (M\"obius strip). And, assuming that a change in topology during neutrino mixing is suppressed by energy ``barriers,'' or other topological ``barriers'' (e.g., one cannot continuously deform a doughnut into a sphere), while neutrino mixing without topology-change is (relatively) enhanced, one can readily explain the experimental observation of (nearly) maximal $\nu_\mu-\nu_\tau$ neutrino mixing---at least maximal $\nu_\mu-\nu_\tau$ mixing over long distances, where the foregoing topological influences would be cumulative [16]. If this explanation is basically correct, then it follows that the neutrino mass-matrix and associated 
mixing-parameters needed to explain bi-maximal neutrino mixing, would be the \emph{result}, at least in part, of these deeper (internal) topological differences between neutrinos, and not their \emph{cause}.

\setcounter{section}{6}
\setcounter{subsection}{-1}

\subsection{Conclusions}

It is widely believed that the explanation for fundamental-fermion (quark and lepton) family replication is to be found in theories of quantum gravity (e.g., superstrings).  And, yet,
as demonstrated here and elsewhere [6], a simple
``analytic continuation''
of a Hermitian matrix representing the fermion-number operator, leads to a
new, and unconventional,
\emph{internal}  description of \emph{quarks} and \emph{leptons}, which also explains family replication. In particular, this
description, 
unlike the conventional standard-model description, is capable of explaining, among
other things, the fact that there are just \emph{three} observed families of quarks and leptons. 
We take these  facts to be evidence that the (phenomenological) ``analytic continuation'' $\Fop\to\Fv$, or at least the result of the continuation $\Fv$, somehow reflects physics at the Planck level where flavor degrees-of-freedom presumably originate. 

It seems that the best chance to show that $\Fv$ and Planck-level physics are related, lies in an appropriate application of superstring theory. Accordingly, the author hopes that the new description of families presented here (see also Refs.\ 6 and 16) will encourage string theorists working on so-called realistic (free-fermionic) three-generation string models (e.g., see Refs.\ 17, 18), to take up the challenge of showing that these models either do, or do not, justify the new description.

\setcounter{section}{7}
\setcounter{subsection}{-1}

\subsection{References and Footnotes}

\quad$\,$ [1] S. Weinberg, \emph{The Quantum Theory of Fields, Vol. I, Foundations}, Cambridge University Press, New York, NY (1995), pp.\ 529--531; \emph{The Quantum Theory of Fields, Vol.\ II, Modern Applications}, Cambridge University Press, New York, NY (1996), p.\ 155.

[2] J.\ M.\ Maldacena, ``Gravity, Particle Physics and Their Unification,'' [hep--ph/0002092].

[3] M.\ B.\ Green, J.\ H.\ Schwarz and E.\ Witten, \emph{Superstring Theory, Vol.\ 1 and 2}, Cambridge University Press, 1987.

[4] N.\ Arkani-Hamed, S.\ Dinopoulos and G.\ Dvali, \emph{Phys.\ Lett., B429}, {\bf 263} (1998) and [hep--ph/9803315].

[5] Ruel V.\ Churchill, \emph{Complex Variables and Applications}, McGraw-Hill Book Company, Inc., New York, NY (1960), pp.\ 259--268. The term ``analytic continuation'' usually refers to an individual analytic function. However, we are dealing here with the ``analytic continuation'' of a 2 by 2 \emph{matrix} $\Fop$ whose components are four different, but closely related, analytic functions ($\pm\cos\theta, \sin\theta e^{\pm i\phi}$). Because the term ``analytic continuation'' has a precise mathematical meaning, and because we would prefer to avoid confusion with well-established mathematical terminology, we will relax this precision in favor of the term ``continuation'' (without quotes) or ``analytic continuation'' (with quotes) whenever we refer to the continuation or ``analytic continuation'' of the matrix $\Fop$.

[6]  Gerald L. Fitzpatrick, \emph{The Family Problem-New Internal Algebraic and Geometric 
Regularities}, Nova Scientific Press, Issaquah, WA (1997).  Additional information: 
http://physicsweb.org/TIPTOP/ or\hfill\break
 http://www.amazon.com/exec/obidos/ISBN=0965569500.
In spite of the many successes of the standard model of particle physics, the observed 
proliferation of matter-fields, in the form of ``replicated" generations or families, is a major 
unsolved problem.  In this book I propose a new organizing principle for fundamental fermions, 
i.e., a minimalistic ``extension" of the standard model based, in part, on the Cayley-Hamilton 
theorem for matrices.   In particular, to introduce (internal) global degrees of freedom that are 
capable of distinguishing all observed flavors, I use the Cayley-Hamilton theorem to generalize 
the familiar standard-model concept of scalar fermion-numbers $f$ (i.e., $f_m=+1$ for all fermions and 
$f_a=-1$ for all antifermions).  This theorem states that \emph{every} (\emph{square}) 
\emph{matrix satisfies its 
characteristic equation}. Hence, if $f_m$ and $f_a$ are taken to be the eigenvalues of some real
 matrix $\Fv$---%
a ``generalized fermion  number"---it follows from this theorem that both $f$ and $\Fv$ are square-%
roots of unity.  Assuming further that the components of both $\Fv$ and its eigenvectors are global 
charge-like quantum observables, and that $\Fv$ ``acts" on a (real) vector 2-space, both the form 
of $\Fv$
and the 2-space metric are determined.  I find that the 2-space has a non-Euclidean or 
``Lorentzian" metric, and that various associated 2-scalars serve as global flavor-defining 
``charges," which can be identified with charges such as strangeness, charm, baryon and lepton 
numbers etc..  Hence, these global charges can be used to describe individual flavors (i.e., flavor 
eigenstates), flavor doublets and families.  Moreover, because of the aforementioned non-%
Euclidean constraints, and certain standard-model constraints, I find that these global charges are 
effectively- ``quantized" in such a way that families are replicated.  Finally, because these same 
constraints dictate that there are only a limited number of values these charges can assume, I find 
that families always come in ``threes."

[7] J.\ Bernstein, \emph{Elementary Particles and Their Currents}, W.\ H.\ Freeman and Co., San Francisco (1968), pp.\ 23--25.

[8] T. D. Lee, \emph{Particle Physics and Introduction to Field Theory
Vol.\ I}, Harwood Academic Publishers, New York, NY (1981), pp.\ 210--211. 

[9] Even though $\Fv_{\diag}=\Fop=\sigma_z$, these two matrices act on entirely different spaces, since $\Fop$ is associated with the \emph{constant} $\cos\theta=1$, and the \emph{variable} phase-factor $e^{-i\phi}$, whereas $\Fv$ is associated with the \emph{variable} $\cos\theta\ge 1$,  and the \emph{constant} phase-factor 
$e^{+
i\phi}=+i$. 

[10]  When weak interactions are ``turned off'' flavor eigenstates and mass 
eigenstates are one and the same.  For the most part, when we speak here of flavor eigenstates, we are referring to the situation where flavor- 
and mass-eigenstates are the same.

[11] Strictly speaking, besides the specification of global charges, the overall quantum state of a fundamental fermion would, necessarily, involve a specification of the spin state, the energy-momentum state and so on, together with a specification of the particular mix of local color (gauge)-charges $R$, $W$, $B$, $G$ and $Y$ carried by each fundamental fermion. This color-mix would be determined, in turn, by a complementary, local $SU(5)$ color-dependent gauge description.

[12] Any acceptable $2\times 2$ matrix $\Fv$ possesses just \emph{two}, \emph{real} linearly-independent eigenvectors, call them $\bQ$ and $\bQ^c$, corresponding to the two \emph{real} eigenvalues $f_m$ and $f_a$, respectively. Therefore, the matrix $\Fv$ can be thought of as ``producing'' the conventional single-particle fermion numbers $f_m$ and $f_a$ via the 2-space eigenvalue equations 
\[
\Fv\bQ = f_m\bQ
\]
and
\[
\Fv\bQ^c=f_a\bQ^c,
\]
respectively.

The 2-vector $\bQ$ (and its scalar components---the \emph{electric} charges of quarks and leptons) describes \emph{matter}, while the linearly-independent 2-vector $\bQ^c$ describes its antimatter counterpart. The superscript $c$ on $\bQ^c$ is merely a label signifying antimatter. It is not an exponent or a symbol for complex conjugation. As such, it signifies only that \emph{the} 2-\emph{vectors $\bQ$ and $\bQ^c$ are real vectors associated with} (``\emph{carried by},'' ``\emph{representing},'' \emph{etc.{\rm)} individual fundamental-fermions or antifermions, respectively}, not state vectors in some Hilbert space. 

Even though the 2-vector $\Q$ $(\Q^c%
)$ does not represent a quantum state, it is associated with the phase factor $e^{i\alpha\,f_m}$ $(e^{i\,\alpha\,f_a})$ associated with a quantum state describing matter (antimatter). To see this, replace $\Fop$ in Eq.\ 1 in the main text by $\Fv$. Then $U$ is replaced by $U'$ where
\[
U'\Q=e^{i\,\alpha\,f_m}\Q
\]
and
\[
U'\Q^c=e^{i\,\alpha\,f_a}\Q^c.
\]
Note that because $e^{i\,n(\delta\alpha)\Fv}= e^{i\,\alpha\Fv}$ when $n$ is very large and $\delta\alpha$ is very small, but nonzero (i.e., $n\,\delta\alpha=\alpha)$, $\alpha$ can be any \emph{finite} number ranging from zero to infinity. Similar arguments apply to the matrix $U$ since $U(\alpha)\cdot U(\alpha')=U(\alpha+\alpha')$.

[13] When we say that the vectors $\bQ$, $\bU$ and $\bV$ are \emph{observables}, we mean that their associated component-``charges'' are mutually-commuting simultaneous observables. Hence, all of these charge-like components can be known in principle, at the \emph{same} time, meaning that the vectors $\bQ$, $\bU$ and $\bV$ can be known simultaneously.

[14] T.\ Kajita, for the Super-Kamiokande, Kamiokande Collaboration, [hep--ex/9810001].

[15] H.\ Georgi and S.\ L.\ Glashow, ``Neutrinos on Earth and in the Heavens,''
 [hep--ph/9808293].

[16] G.\ L.\ Fitzpatrick, ``Topological Constraints on Long-Distance Neutrino Mixtures,'' [aps1999feb12\underbar{~}001] 
available at:
http://publish.aps.org/eprint/

[17] A.\ E.\ Faraggi, ``Towards the Classification of the Realistic Free Fermion Models,'' 
[hep-th/9708112].

[18] G.\ B.\ Cleaver, A.\ E.\ Faraggi, D.\ V.\ Nanopoulos and T.\ ter Veldhuis, ``Towards String Predictions,'' [hep-ph/0002292].

\end{document}